\documentclass[aps,prl,floatfix,showpacs,preprintnumbers,amsymb,a4paper,superscriptaddress,twocolumn
]{revtex4-1}
\usepackage{amsmath}
\usepackage{graphicx}
\usepackage{amssymb}
\usepackage{amsmath}
\usepackage{epstopdf}
\usepackage{float}
\usepackage{color}

\newcommand{\cev}[1]{\reflectbox{\ensuremath{\vec{\reflectbox{\ensuremath{#1}}}}}}
\newcommand{\be}{\begin{equation}}
\newcommand{\ee}{\end{equation}}
\newcommand{\ket}[1]{\left|{#1}\right\rangle}
\newcommand{\bra}[1]{\left\langle{#1}\right|}

\newcommand{\bkt}[2]{\langle{#1}|{#2}\rangle}

\begin{document}
\title{A Hamiltonian for the inclusion of spin effects in long-range Rydberg molecules}
\author{Matthew T. Eiles}
\affiliation{Department of Physics and Astronomy, Purdue University, West Lafayette, Indiana 47907, USA.}
\affiliation{Kavli Institute of Theoretical Physics, University of California, Santa Barbara, Santa Barbara, California 93106, USA}
\author{Chris H. Greene}
\affiliation{Department of Physics and Astronomy, Purdue University, West Lafayette, Indiana 47907, USA.}
\affiliation{Kavli Institute of Theoretical Physics, University of California, Santa Barbara, Santa Barbara, California 93106, USA}
\affiliation{ Purdue Quantum Center, Purdue University, West Lafayette, Indiana 47907, USA.}

\date{\today }

\begin{abstract}
The interaction between a Rydberg electron and a neutral atom situated inside its extended orbit is described via contact interactions for each atom-electron scattering channel. In ultracold environments, these interactions lead to long-range molecules with binding energies typically ranging from $10$-$10^4$MHz. These energies are comparable to the relativistic and hyperfine structure of the separate atomic components. Studies of molecular formation aiming to reproduce observations with spectroscopic accuracy must therefore include the hyperfine splitting of the neutral atom and the spin-orbit (SO) splittings of both the Rydberg atom and the electron-atom interaction.   Adiabatic potential energy curves (PECs) and permanent electric dipole moments (PEDMs) are presented for Rb$_2$ and Cs$_2$. The influence of spin degrees of freedom on the PECs and multipole moments probed in recent experimental work is elucidated, and the observed dipole moments of butterfly molecules are explained by the generalized $^3P_J$-pseudopotential derived here.
\end{abstract}

\maketitle

\begin{section}{Introduction}
 Rydberg atoms, owing to their exaggerated properties, provide a pristine environment for highly accurate quantum metrology and manipulation, especially  in conjunction with the remarkable precision afforded by ultracold laboratory systems and high resolution laser spectroscopy. They reveal a wealth of information about the myriad effects of external fields on quantum systems, the transition between quantum and classical physics, and the universal properties of many different atomic species \cite{Harmin,Zoller,Wunner, Stroud, Zoller2,RMP}. Additionally, they provide a promising framework for quantum information due to the Rydberg blockade and associated long-range interactions \cite{Saffman,Molmer}. Rydberg atoms embedded in dense gases serve as probes of the electron-atom scattering properties of the surrounding gas \cite{Fermi}, and, remarkably, can even bind one or more nearby atoms into an ultra-long-range molecule through the localized interaction induced by this scattering \cite{GreeneSadeghpourDickinson}. The binding energies of these fragile molecules are similar to the relativistic spin-orbit (SO) corrections to the Rydberg atom's level structure, the SO splitting of the electron-atom scattering channels, and the hyperfine splitting of the perturbing atom. These quantities have been measured to increasingly high accuracy in recent years \cite{Gallagher2016a,Gallagher2016b,CsHF}. The improved knowledge of these parameters provides the necessary ingredients to characterize the properties of exotic Rydberg molecules to very high accuracy, provided a full theoretical model including all these effects is developed.
  
The original theoretical predictions of these molecules focused on the simplest cases of $^3S$ \cite{GreeneSadeghpourDickinson} and $^3P$  \cite{HamiltonGreeneSadeghpour,KhuskivadzeJPB} scattering of electrons by a Rb atom, neglecting hyperfine structure and the $^3P_J$ splittings. Essentially simultaneously, sophisticated Green's function approaches for the same system were developed \cite{HamiltonThesis,Crowell,KhuskivadzeJPB}; one in particular (henceforth referenced as KCF) included the $^3P_J$ splitting \cite{KhuskivadzeChibisovFabrikant}. Photoassociation experiments have since formed $ns$, $np$, and $nd$ Rydberg molecules of both Rb and Cs \cite{Bendkowsky,quantumreflection,Tallant,Sass,Pfau,AndersonPRL}, and recently even ``trilobite'' and ``butterfly'' molecules consisting of large admixtures of high-$l$ states with very large permanent electric dipole moments (PEDMs) \cite{trilobite,butterfly}.    Several recent works have explored effects related to the hyperfine splitting in these molecules, beginning with a joint theoretical and experimental investigation \cite{AndersonPRA,AndersonPRL} and recently including studies of mixed singlet-triplet potentials \cite{SingTripMix,Sass,Markson} and the ability of this mixing to induce a spin-flip \cite{Niederprum}. 

The present article expands upon this wide body of literature, particularly the theoretical efforts \cite{KhuskivadzeChibisovFabrikant,AndersonPRA,Markson}, by including all relevant interactions for both alkali atomic species of common experimental interest, thus unifying past theoretical approaches into a complete model for the first time. A major component of this is the inclusion of $^3P_J$ splittings within the pseudopotential approach, which is important for quantitative calculations with heavier atoms like Rb and Cs. This provides highly accurate potential energy curves (PECs), which will be utilized in a forthcoming effort to fully confirm and understand past experimental observations \cite{paper2}; already in the present effort the $^3P_J$ splitting gives improved
theoretical values for the observed dipole moments of butterfly molecules.  These potential curves additionally reveal a wealth of interesting regimes for future experimental and theoretical exploration. An improved understanding of these exotic molecules provides insight into the precision study of the scattering properties and control possibilities of these molecules, and provides a strong foundation for studies of many-body and mean-field effects in polyatomic systems, an area of current interest that demands accurate two-body information \cite{Demler,Schlag,FeyKurz,EilesJPB, MolSpec,RydbergRev}. Additionally, the modified $^3P_J$-wave pseudopotential could find applications in other ultracold systems or in parallel systems in nuclear physics \cite{Omont,Idziaszek,Nuke1,Nuke2}. Note that the present treatment, which aims to replace the Green's function technique by a phase shift-dependent operator that can be numerically diagonalized, connects with the spirit and motivations of effective field theory \cite{LePage}.
\section{construction of the hamiltonian matrix}
\begin{figure}[tbp]
{\normalsize 
\includegraphics[scale =0.5]{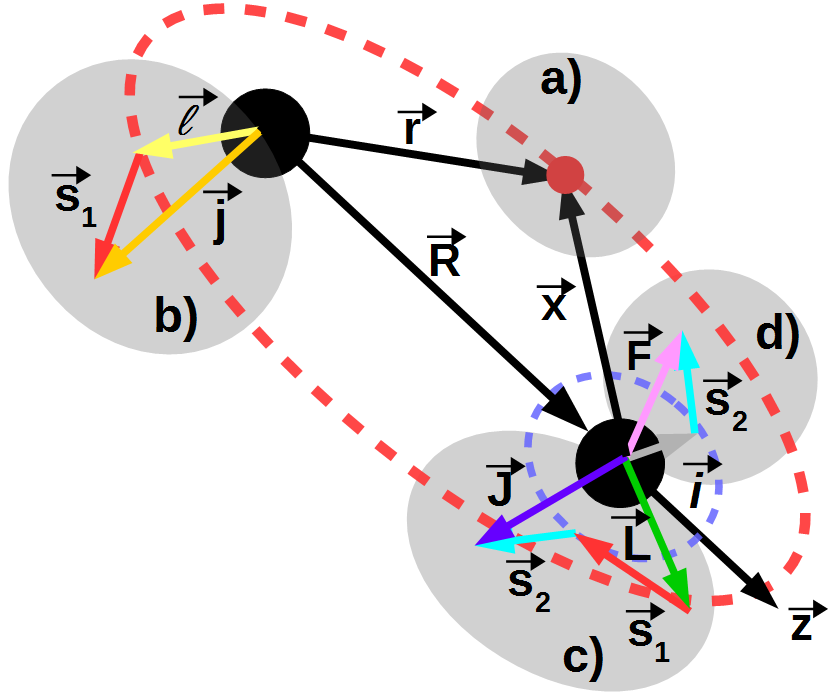}
}
\caption{The molecular system and relevant angular momenta. The internuclear axis lies parallel to the body-frame $z$ axis passing through the ionic core (left) and the ground state atom (right). The red (blue) dashed oval represents the Rydberg (ground state) electron's orbit. a) The Rydberg electron is located at $\vec r$ relative to the core and at $\vec X = \vec r - \vec R$ relative to the perturber.  b) The spin of the Rydberg electron, $\vec s_1$ (red), couples to its orbital angular momentum relative to the core, $\vec l$ (yellow), to give a total angular momentum $\vec j$ (orange) with projection $m_j = m_l + m_1$. c) The interaction between the Rydberg electron and neutral atom depends on the total electronic spin, $\vec S = \vec s_1$ (red) $+ \vec s_2$ (cyan), coupled to the orbital angular momentum $\vec L$ (green) relative to the perturber to form total angular momentum $\vec J$ (purple), with projection $M_J = m_l+m_{1} + m_{2}$. 
d) The spin of the perturber's outer electron, $\vec s_2$ (cyan) interacts with the perturber's nuclear spin, $\vec i$ (gray) to form $\vec F$ (pink) and its projection $M_{F} = m_{2} + m_i$. The only good quantum number of the combined system is $\Omega = m_j+m_{2} + m_i$. }
\label{fig:cartoon}
\end{figure}

A diatomic system of alkali atoms, one in its ground state and the other in a highly excited Rydberg state, is considered.  The two nuclei adiabatically traverse PECs $v_i(\vec R)$ that, in the Born-Oppenheimer approximation, depend parametrically on the internuclear distance, $\vec R = R\hat z$. The much faster electronic motion defines these PECs through the time-independent Schr\"{o}dinger equation, $H(\vec r;\vec R)\Psi_i(\vec r;\vec R)=v_i(\vec R)\Psi_i(\vec r;\vec R)$ for the electronic wave function $\Psi_i(\vec r;\vec R)$. This equation is solved by diagonalizing the matrix representation of $H(\vec r;\vec R)$. This method is chosen in contrast to alternative methods based on the Coulomb Green's function to simplify the inclusion of spin degrees of freedom. Additionally, diagonalization immediately provides the eigenfunctions, yielding multipole moments, information about state mixing, and non-adiabatic couplings.  The full Hamiltonian includes all relevant relativistic effects:
\be
\label{eq:Hamiltonian}
\hat H(\vec r;\vec R) = \hat H_{Ryd}(\vec r) + \hat V_P(\vec R,\vec r) + \hat H_{HF} - \frac{\alpha}{2R^4}.
\ee
The Hamiltonian of the Rydberg atom,  $\hat H_{Ryd}(\vec r)$,  includes the effects of core electrons and the Rydberg SO splitting, typically parameterized by measured quantum defects from atomic spectroscopy.  $\hat V_P(\vec R,\vec r)$ is the electron-perturber pseudopotential generalized to include all electron-scattering channels up to $P$-wave: $^1S_0$,$^3S_1$,$^1P_1$, and $^3P_{0,1,2}$. $\hat H_{HF}$ is the hyperfine interaction between the perturber's nuclear and electronic spins, and $-\frac{\alpha}{2R^4}$ is the polarization potential between the Rydberg core ion and the perturber. These terms will be described in more detail as their matrix elements are constructed.
   
Fig. \ref{fig:cartoon} schematizes these different interactions and illustrates the two centers inherent to this system, which are crucial when dealing with the $^3P_J$ scattering states. The first center, the Rydberg ion, determines the good quantum numbers of the Rydberg electron's wave function in the absence of a perturbing atom,  $\ket{n(ls_1)jm_j}$. Explicitly, these are the principal quantum number $n$, the total angular momentum $\vec j^2 = (\vec s_1 + \vec l)^2$, and its projection onto the internuclear axis $m_j$. Since these eigenfunctions are known, a sensible choice of basis to represent the Hamiltonian includes these unperturbed eigenfunctions $\ket{n(ls_1)jm_j}$ as well as the uncoupled nuclear and electronic spin states of the perturber, $\ket{s_2m_2;im_i}$. Diagonalization of $\hat H_{Ryd}(\vec r)$, which would otherwise involve the numerical solution of the electron's dynamics in some model potential describing the alkali atom, is trivial in this basis, with eigenenergies given by experimentally determined quantum defects:
\be
\label{eq:Rydbergformula}
E_{n(s_1l_1)jm} = -\frac{1}{2(n-\mu_{(s_1l_1)j}(n))^2}.
\ee
The quantum defects are parametrized:
\be
\label{eq:quantumdefects}
\mu_{(s_1l_1)j}(n) = \mu_{(s_1l_1)j}(0) + \frac{\mu'_{(s_1l_1)j}(0)}{\left[n - \mu_{(s_1l_1)j}(0)\right]^2}.
\ee
Table \ref{tab:datatable} displays these parameters for $ns$, $np$, $nd$, and $nf$ states of both Rb \cite{LiRb,JamilRb} and Cs \cite{GoyCs,MoreCs}. For higher angular momenta, the quantum defects account for core polarization through the approximate formula
 \begin{align}
 \mu_l(n) &= \left(\frac{\alpha(X^+)[3n^2 - l(l+1)]/4}{n^2(l-1/2)l(l+1/2)(l+1)(l+3/2)}\right),\label{eq:coredefects}
  \end{align}
  where $\alpha(X^+)$ is the polarizability of the Rydberg core of atom $X$.
  Table 1 shows the polarizabilities of both atoms and their positive ions \cite{Polarizabilities}.
  The hydrogenic fine structure splitting is assumed for these nonpenetrating high-$l$ ($l>3$) states:
  \be
  \Delta E_{n(s_1l_1)jm} = -\frac{\alpha^2}{2n^3}\left(\frac{1}{j+1/2} - \frac{3}{4n}\right),
  \ee
  where $\alpha$ is the fine structure constant. 
  Since this splitting and the core polarization-induced quantum defects decrease rapidly with $l$, the $l>2$ states are nearly degenerate and only slightly modify the potential curves. The eigenfunctions of $\hat H_{Ryd}(\vec r)$ are
\be
\label{eq:jdepefuncs}
\psi _{n(ls_1)jm_j}(\vec r) = \sum_{m,m_1}C_{lm,s_1m_1}^{jm_j}\frac{f_{nlj}(r)}{r}Y_{lm}(\hat r)\chi_{m_1}^{s_1},
\ee 
where $\chi_{m_1}^{s_1}$ is the Rydberg electron's spin wave function. 
The radial eigenstates, $f_{n(ls_1)j}(r)$, for low-$l$ states with non-integral quantum defects are approximately given by Whittaker functions. This is an excellent approximation beyond distances of a few Bohr radii where the non-Coulombic potential due to the actual distribution of core electrons vanishes.

\begin{center}
\begin{table}[t]
\begin{tabular}{|| c|| c c| c|| c c||}
\hline
 {\bf Rb} & $\mu(0)$ & $\mu'(0)$ &{\bf Cs} & $\mu(0)$ & $\mu'(0)$ \\ 
  \hline \hline
  $s_{1/2}$& 3.1311804 & 0.1784 & $s_{1/2}$& 4.049325 & 0.2462   \\
 $p_{1/2}$ & 2.6548849 & 0.2900 & $p_{1/2}$& 3.591556 & 0.3714   \\
 $p_{3/2}$ & 2.6416737 & 0.2950 & $p_{3/2}$& 3.559058 & 0.374  \\
 $d_{3/2}$ & 1.34809171 & -0.60286& $d_{3/2}$& 2.475365 & 0.5554  \\
 $d_{5/2}$ & 1.34646572 & -0.59600& $d_{5/2}$& 2.466210 & 0.067  \\
 $f_{5/2}$ & 0.0165192 & -0.085 & $f_{5/2}$& 0.033392 & -0.191   \\
 $f_{7/2}$ & 0.0165437 & -0.086 & $f_{7/2}$& 0.033537 & -0.191   \\
 \hline \hline
  & Rb & Rb$^+$ &   & Cs  & Cs$^+$\\
  \hline
  $\alpha$ (a.u.) & 319.2 & 9.11 & $\alpha$ & 402.2 & 15.8\\
  \hline \hline
   & Rb(ns) & Rb(5s) &  & Cs(ns) & Cs(6s)  \\
   \hline
   A (GHz) & $18.55/(n^*)^3$ & 3.417  & A & $3.383/(n^*)^3$ & 2.298 \\
 \hline
\end{tabular}
\caption{Quantum defects, polarizabilities $\alpha$, and hyperfine constants $A$ for $^{87}$Rb and $^{133}$Cs from \cite{LiRb,JamilRb,GoyCs,MoreCs,CsHF,Polarizabilities,RbHF,AllHF}}.
 \label{tab:datatable}
\end{table}
 \end{center}

The hyperfine Hamiltonian is $\hat H_{HF}=A\vec I\cdot \vec S_2$. Table \ref{tab:datatable} gives the constant $A$; for the Rydberg atom this decreases as $n^{-3}$ and is irrelevant at the level of accuracy considered here \cite{CsHF,RbHF,AllHF}.  The matrix elements of $\hat H_{HF}$ in the uncoupled basis are
    \begin{align}
  &\bra{\alpha im_i,s_2m_2}A\vec I\cdot\vec S_2 \ket{\alpha' i'm_i',s_2'm_2'}= \frac{A}{2}\delta_{\alpha\alpha'}\sum_{FM_F}C_{s_2m_2,im_i}^{FM_F}\nonumber\\&\times C_{s_2m_2',im_i'}^{FM_F}\left[(F(F+1)-i(i+1)-s_2(s_2+1)\right],
  \end{align}
  where $\alpha = \{n,l,s_1,j,m_j\}$ and the nuclear spin  $i= 3/2(7/2)$ for $^{87}$Rb($^{133}$Cs). $C_{j_1m_1,j_2m_2}^{j_3m_3}$ is a Clebsch-Gordan coefficient.
  
 In the context of Rydberg physics, the pseudopotential $\hat V_P(\vec R,\vec r)$ determining the interaction between the electron and the perturber was first derived for the $S$ partial wave by Fermi, and then generalized to $P$-wave by Omont \cite{Fermi,Omont}. It has been verified and used in a variety of other physical contexts \cite{DuGreene89,Idziaszek,Lebedev}. A contact potential is justified based on the large wavelength of the electron relative to the size of the perturber, motivating a partial wave expansion of the Rydberg electron wave function relative to the perturbing atom. To incorporate the SO splitting of this scattering process these partial waves $(L)$ are coupled with the total electron spin $(S)$ to give phase shifts depending on the total angular momentum $J$ of the Rydberg electron about the perturber.  Khuskivadze {\it et al.} \cite{KhuskivadzeChibisovFabrikant} were the first to include this via a Green's function treatment and a finite range potential for the electron-atom interaction.  Here we develop an alternative procedure that is much more convenient for the diagonalization treatments with zero-range interactions that are typically implemented. The small coupling between $^1P_1$ and $^3P_1$ symmetries is neglected so that the Fermi pseudopotential remains diagonal in spin.
 
 Two complementary approaches are considered here to emphasize different aspects of this derivation and provide alternative physical pictures. The first involves a recoupling of the tensorial operators involved in the Fermi pseudopotential so that $J$-dependent phase shifts can be incorporated, while the second reformulates the pseudopotential so that it is diagonal in the $\ket{(LS)J\Omega}$ basis with matrix elements proportional to the tangents of the $J$-dependent phase shifts, and then considers an expansion of the electronic wave function near the perturber. The first approach begins with the Fermi pseudopotential including singlet/triplet states:
\begin{equation}
\label{eqn:recouplingppstate}
\hat V=\mathcal{A}(SL,k) \sum_{M_{S}}\chi^S_{M_S}(\chi^S_{M_S})^\dag\cev\nabla {}^{L }\delta(\vec X)\cdot \vec\nabla^L.
\end{equation}
Here $\vec X = \vec r - \vec R$, and  $\mathcal{A}(SL,k) = (2L+1)2\pi a(SL,k)$, where $a(SL,k)$ is the energy dependent scattering length(volume) for $L= 0(1)$ and for $S=0$ or $1$. $\cev\nabla$ and $\chi^S_{M_S} $ represent conjugate operators acting to the left. Eq. \ref{eqn:recouplingppstate} is expressed using zero-rank tensor operators composed of the tensorial sets $\chi_{M_S}^{(S)}$ and $\nabla_{M_L}^{(L)}$ via standard angular momentum theory:
 \begin{align*}
\hat V &=  \mathcal{A}(SL,k)\delta(\vec X)\sqrt{(2L+1)(2S+1)}(-1)^{-L-S} \\&\times \left\{\left[\cev{\nabla}{}^{(L)}\times\vec\nabla^{(L)}\right]^{(0)}\times \left[ \chi^{(S)}\times(\chi^{(S)})^\dag\right]^{(0)}\right\}^{(0)}_0 .
 \end{align*}
The $J$-dependence is included by recoupling these operators in the usual spirit of Wigner-Racah algebra \cite{FanoRacah,
FanoMacekRMP,GreeneZareAnnRev}. The recoupling coefficient is calculated using properties of Wigner 9J symbols, and $\mathcal{A}(SLJ,k)$ may now be brought inside the final scalar product and allowed to become $J$-dependent:
\begin{align}
\hat V &= \delta(\vec X)\sum_J\mathcal{A}(SLJ,k)\sqrt{2J+1}(-1)^{-L-S}\nonumber\\& \times\left\{\left[\cev\nabla{}^{(L)}\times\chi^{(S)}\right]^{(J)}\times \left[\vec\nabla^{(L)}\times (\chi ^{(S)})^\dag\right]^{(J)}\right\}_0^{(0)}\nonumber.\nonumber
\end{align}
After decoupling, this scalar operator is
 \begin{align}
 \label{eqn:approachonefinal}
   \hat V&= \delta(\vec X)\sum_J \sum_\Omega\sum_{M_L,M_L'} \mathcal{A}(SLJ,k) C_{LM_L,S\Omega-M_L}^{J\Omega}\\&\times C_{LM_L',S\Omega - M_L'}^{J\Omega}\cev\nabla{}^{(L)}_{M_L}(\chi ^{(S)}_{\Omega-M_L})^\dag\vec\nabla^{(L)}_{M_L'} \chi^{(S)}_{\Omega-M_L'}.\nonumber
\end{align}
After uncoupling the Rydberg electron's spin and orbital angular momenta, then coupling the electronic spins together, matrix elements of $\hat V$ in the basis chosen above are constructed using the definition
 \be
  Q_{LM_L}^{nlj}(R) =\delta_{m,M_L} \left[\vec \nabla^L\left(\phi_{nljm}(\vec R)\right)\right]_{M_L}^L,\nonumber
   \ee
   where
   \begin{align}
\label{eqn:Qfuncs}
Q_{00}^{nlj}(R)&= \frac{f_{nlj}(R)}{R}\sqrt{\frac{2l+1}{4\pi}},\\
Q_{10}^{nlj}(R) &= \sqrt{\frac{2l+1}{4\pi}}\partial_R\left(\frac{f_{nlj}(R)}{R}\right),\\
Q_{1\pm 1}^{nlj}(R)&=\frac{f_{nlj}(R)}{R^2}\sqrt{\frac{(2l+1)(l+1)l}{8\pi}},l>0.
\end{align}  
The pseudopotential matrix $\underline{V}$ which results is given in Eq. (\ref{eqn:scattpotential}), which we now derive directly following the alternative second approach. This starts from a reformulation of the Fermi pseudopotential in which all the angular dependence has been projected out. This explicitly incorporates $J$-dependent scattering phase shifts by projecting into states with good quantum numbers $\beta=\left\{(LS)J\Omega\right\}$ describing the electron-atom interaction:
\begin{align}
\label{finalprojectionform}
\hat V_P&=\sum_{\beta}\ket{\beta}\frac{(2L+1)^2}{2}a(SLJ,k)\frac{\delta(X)}{X^{2(L+1)}} \bra{\beta}.
\end{align}
 Here,
\be\langle\hat X\ket\beta = \sum_{M_L,M_S}C_{LM_L,SM_S}^{JM_J}Y_{LM_L}(\hat X)\chi_{M_S}^S.\ee
 Further details explaining the integration of the angular terms of the gradient in Eq. (\ref{finalprojectionform}) are found in appendix A.  Since the good quantum numbers $\beta$ are incompatible with those characterizing the eigenstates of the Rydberg electron, the Rydberg wave function of Eq. (\ref{eq:jdepefuncs}) is expanded to first order about the position of the perturber:
 \begin{align}
 &\psi_{n(ls_1)jm_j}(\vec r) =\\ &\sum_{m,m_1}C_{lm,s_1m_1}^{jm_j}\chi_{m_1}^{s_1}\left[\phi_{nljm}(\vec R) + \vec\nabla\left(\phi_{nljm}(\vec R)\right)\cdot\vec X\right],\nonumber
 \end{align}
 where $\phi_{nljm}(\vec R) = \frac{f_{nlj}(R)}{R}Y_{lm}(\hat R)$. 
   After using the spherical tensor representation of $\vec\nabla\phi_{nljm}(\vec R)$ given by the $Q$ functions and expressing $\vec X$ in terms of spherical harmonics $Y_{LM}(\hat X)$ centered at the perturber, it becomes clear that this expansion mediates the transformation from spherical harmonics relative to the Rydberg atom, $Y_{lm}(\hat r)$, to $S$ and $P$ partial waves relative to the perturber, $Y_{LM}(\hat X)$:
\begin{align}
  \label{taylorexpform}
  \psi_{n(ls_1)jm_j}(\vec r)&= \sum_{m_1=-s_1}^{m_1=s_1}\sum_{L=0}^1\sum_{M_L=-L}^{M_L=L}X^Lf_L\\& \times C_{lM_L,s_1m_1}^{jm_j}Q_{LM_L}^{nlj}(R)Y_{LM_L}(\hat X)\chi_{m_1}^{s_1},\nonumber
  \end{align}
  where $f_L = \sqrt{\frac{4\pi}{(2L+1)}}$. 
  Coupling $\psi_{n(ls_1)jm_j}(\vec r)$ from Eq. (\ref{taylorexpform}) to the perturber's spin introduces $S=0,1$ states:
\begin{align}
\label{singtripform}
 &\psi_{n(ls_1)jm_j}(\vec r)\chi_{m_2}^{S_2}= \sum_{m_1=-s_1}^{m_1=s_1}\sum_{\substack{L=0\\M_L=-L}}^{\substack{L=1\\M_L=L}}\nonumber\sum_{\substack{S=0\\M_S=-S}}^{\substack{S=1\\M_S=S}}X^L\chi_{M_S}^S\\&\times C_{lM_L,s_1m_1}^{jm_j}C_{s_1m_1,s_2m_2}^{SM_S} Q_{LM_L}^{nlj}(R)f_LY_{LM_L}(\hat X).
  \end{align}
The matrix elements of this operator are obtained from Eq. \ref{singtripform} after a trivial integration over $X$ and introducing the Clebsch-Gordan coefficients $C_{LM_L,SM_S}^{J\Omega}=\bkt{(LS)J\Omega}{LM_L,SM_S}$. These matrix elements are compactly expressed by first constructing the matrix representation of Eq. (\ref{finalprojectionform}) in the basis of quantum numbers centered at the perturber, $\ket\beta=\ket{(LS)J\Omega}$:
  \begin{align}
  \label{eqn:diagpotential}
U_{\beta,\beta'} &= \delta_{\beta,\beta'}\frac{(2L+1)^2} {2} a(SLJ,k).
  \end{align}
The transformation of this diagonal matrix into one in the $\ket{\alpha s_2m_2}$ basis, where $\alpha = \{n,l,s_1,j,m_j\}$, is mediated by a ``frame-transformation'' matrix $\mathcal{A}$. In this context this is physically equivalent to a change of coordinates and good quantum numbers between the two geometrical centers of this system, analogous to what is done in multiple scattering theory \cite{DillDehmerJCP}. This matrix is readily deduced from the prior steps of the derivation:
  \begin{align}
 \mathcal{A}_{\alpha s_2m_2,\beta}  &  =\sum_{M_L=-L}^{M_L=L}f_{L} C_{lM_L,s_1m_j-M_L}^{jm_j}Q_{LM_L}^{nlj}(R)\\&\,\,\,\,\times C_{s_1m_j-M_L,s_2m_2}^{Sm_j-M_L+m_2}C_{LM_L,Sm_j-M_L+m_2} ^{Jm_j+m_2}. \nonumber
  \end{align}
The final scattering matrix is diagonal in $m_i$ and for every $n$ and $l$ consists of a block matrix:
 \begin{align}
 \label{eqn:scattpotential}
 &\underline{V}=\underline{\mathcal A}\times\underline{U}\times\underline{\mathcal A^\dagger}.
 \end{align}
 These matrix elements can be equivalently obtained from Eq. \ref{eqn:approachonefinal} after the same recoupling of the basis states, but without the need for an expansion of the wave function.  The mixing of $M_L,M_L'$ implied by Eqs. \ref{eqn:approachonefinal} and \ref{eqn:scattpotential} is critical for an accurate physical description of this splitting, since the total spin vector $\vec{S}$ and total
orbital $\vec{L}$ precess during each
$P$-wave electron-perturber collision. This was first recognized and incorporated in the Green's function calculation of KCF \cite{KhuskivadzeChibisovFabrikant}. However, all subsequent work has neglected this detail. We expect that the much simpler description developed here using zero-range potentials will correct this oversight. This mixing of $M_L$ projections invalidates the use of $\Sigma$ and $\Pi$ symmetry labels to categorize the $^3P_J$ potential curves. Incidentally, the Clebsch-Gordan coefficients vanish for $M_L = 0$ for the $^3P_1$ state, so that it remains a $\Pi$ state in the absence of the hyperfine interaction.   Appendix B provides more details about this potential matrix without the obscuring complexities of the fine and hyperfine structure. 
  \end{section}
  \begin{section}{Details of the calculation}
  The energy-dependent scattering length for $S$-wave scattering, $a(S0J,k) =- \frac{\tan\delta(0,S,J,k)}{k}$, and the energy-dependent scattering volume for $P$-wave scattering, $a(S1J,k) = -\frac{\tan\delta(1,S,J,k)}{k^3}$, are calculated using the phase shifts of KCF and the semiclassical electronic momentum  $k(R) = \sqrt{\frac{2}{R} - \frac{1}{n_H^2}}$.  $n_H$ is the principal quantum number of the nearest hydrogenic manifold.  The  $^3P_J$ phase shifts for Cs were slightly shifted (by $\sim 1$ meV) from the values calculated by KCF to align their resonance positions with experimental values \cite{pwaveresonance1,pwaveresonance2}. These phase shifts are plotted in appendix C. No direct experimental measurements of the Rb resonance positions yet exist, although an average value consistent with the phase shifts of KCF was extracted from observations of Rb$_2$ Rydberg molecules \cite{quantumreflection}. At very low energies the $S$-wave phase shifts for both species were smoothly connected to experimentally determined zero-energy scattering lengths \cite{Sass,quantumreflection}.

The Hamiltonian matrix $\underline{H}$ is diagonalized at every value of the internuclear distance, $R$. The dimension of this matrix is finite in the spin quantum numbers, while the infinite number of states of different $n$ must be truncated. Typically four total manifolds $\{n_H-2,n_H-1,n_H,n_H+1\}$ are employed in the results presented here. The only good quantum number of this system is the total spin projection, $\Omega = m_j+ m_2 + m_i$. At long-range, where the perturber-electron interaction vanishes, the potential curves can be identified asymptotically via the electronic angular momenta $l$ and $j$, and the perturber's total nuclear spin $F$.  Since only $L\le 1$ partial waves are included in the electron-perturber scattering, only states with $|m_j|\le 3/2$ will be shifted, and so $\underline{H}$ is block diagonal in $\Omega$, $|\Omega|<\frac{7}{2}\left(\frac{11}{2}\right)$ for Rb(Cs).  For states around $n_H = 30$ the basis size ranges from approximately 2200(2000) for Cs(Rb) with $|\Omega|=1/2$, down to 275 for the maximal $\Omega$.
  \begin{figure}[tbp]
{\normalsize 
\includegraphics[scale =0.09]{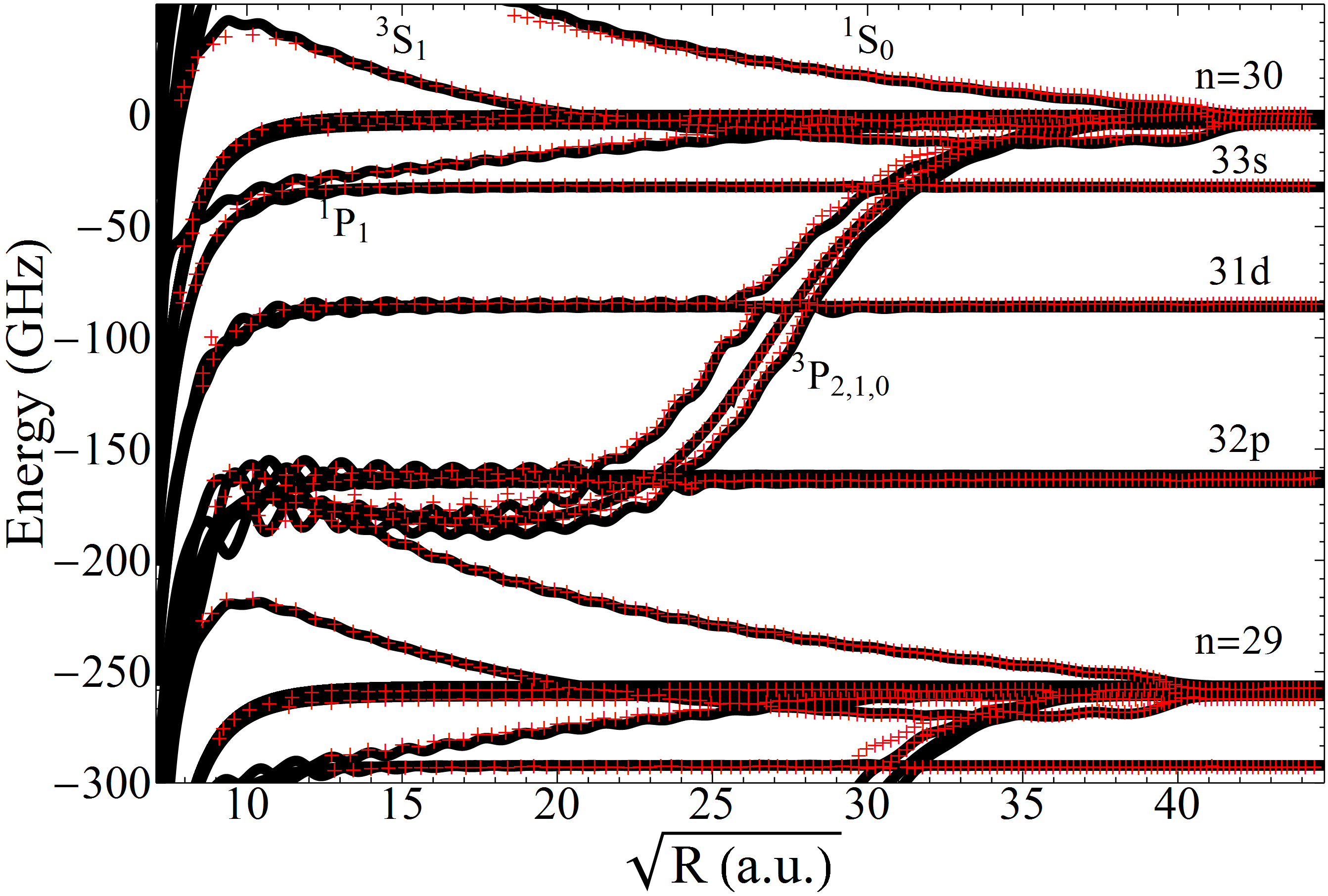}
}
\caption{PECs of Rb$_2$, $\Omega = 0$ (black) without the hyperfine splitting. The results of KCF (red crosses) is also plotted. The abscissa is the square root of $R$, which more uniformly spaces the potential wells. The detuning is relative to $n_H=30$.}
\label{fig:RbM0n30}
\end{figure}
\begin{figure}[h]
{\normalsize 
\includegraphics[scale =0.09]{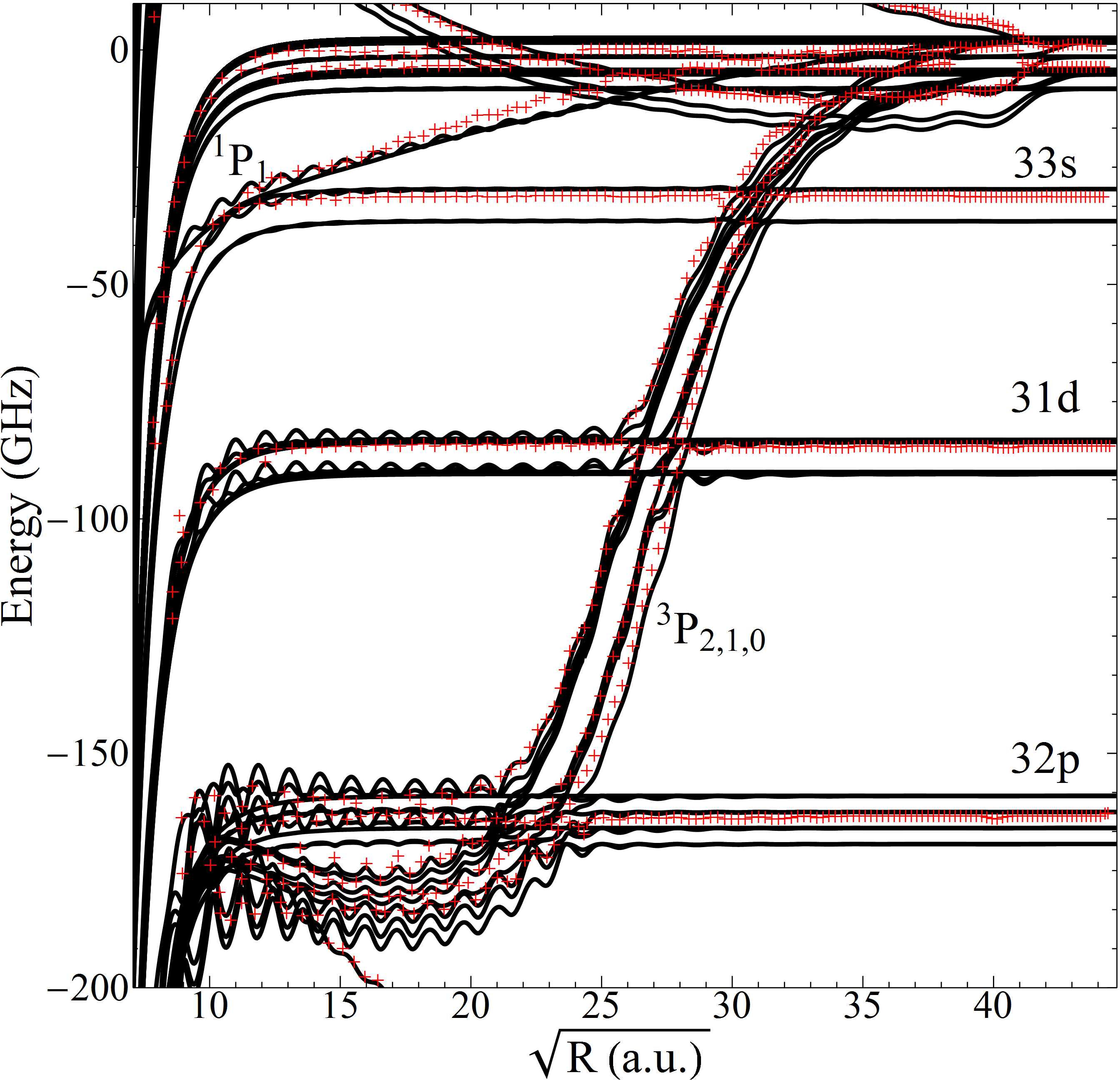}
}
\caption{PECs of Rb$_2$, $\Omega = 1/2$, with the hyperfine splitting of the ground state atom (black). The results of KCF (red crosses) is plotted, although these ignore hyperfine and fine structure splittings. The inclusion of the additional fine and hyperfine structure creates a multitude of additional $^3P_J$-scattered states and splits the trilobite PECs into separate hyperfine states. The detuning is relative to $n_H=30$, and the potential curves are labeled as in Fig. \ref{fig:RbM0n30}.}
\label{fig:RbM1n30}
\end{figure}
The accuracy and convergence of these PECs is a controversial issue.  A number of adjacent manifolds must be included in the basis so that level repulsion constrains the divergences in the scattering volumes caused by the $^3P_J$ shape resonances \cite{HamiltonGreeneSadeghpour}. However, a study of the $ns$ potential wells has shown that the inclusion of additional manifolds deepens these long-range wells uncontrollably due to the highly singular delta function potential \cite{Fey}; numerical tests also show that the deepest butterfly potential wells are sensitive to the basis size (see the discussion of Fig. \ref{fig:convergencecomparison}).   Two independent benchmarks are employed here to find the most satisfactory values for the potential curves, given their formal non-convergence. The Borodin and Kazansky model \cite{BKmodel} (BK hereafter) uses the phase shifts to determine the smooth large-scale structure of the trilobite and butterfly PECs through
\be
\label{eq:BKmodel}
E_{(LS)J}(R)=-\left[2\left(n-\delta(L,S,J,k[R])/\pi\right)^2\right]^{-1}.
\ee This serves as a crude convergence benchmark, since the true PECs should not differ dramatically from these results. The second convergence check is the comparison between the potential curves from the present model with those calculated in KCF. Good agreement with these two benchmarks was found after including one more manifold below the level of interest than above;  specifically, the set $\{n_H -2...n_H+1\}$ is used. The $n^{-3}$ scaling of the Rydberg level spacing lends some physical justification to this heuristic approach, since the manifolds above the level of interest contribute more weight to the level repulsion due to their relative closeness in energy; the additional manifolds below ``balance'' this repulsion.  For clarity, only comparisons with KCF and not the BK comparisons are included in the figures.  Some further nuances and convergence tests will be discussed in later sections. 
 \end{section}

  \begin{section}{Adiabatic potential energy curves}

As a straightforward confirmation of the validity of this full theory, the PECs for Rydberg energies around $n_H=30$ are compared with the calculations of KCF. Figs. \ref{fig:RbM0n30}-\ref{fig:CsM1n30} show these comparisons and reveal a wealth of information. In Fig. \ref{fig:RbM0n30}, the hyperfine structure is neglected for clarity. The main features of KCF are reproduced excellently, validating this basis set truncation and the accuracy of our $^3P_J$ pseudopotentials.   Low-$l$ molecules can be adequately described without the $^3P_J$ splitting, since the butterfly potentials cross the low-$l$ states with comparable slopes and distances, although quantitative results still require this level of accuracy. The $J$-dependence become qualitatively crucial in the depths of the butterfly states and in their PEDMs (see Figs. \ref{fig:Rbbutterfly},\ref{fig:dipoles}).

Inclusion of the hyperfine structure adds significant complexity: it increases the multiplicity of butterfly states, further mixes these states, introduces many avoided crossings, and splits the low-$l$ states by several GHz. Fig. \ref{fig:RbM1n30} shows results for Rb$_2$ with $n_H\sim 30$  and $\Omega = 1/2$, highlighting the importance of these additional splittings in shifting the long-range asymptotes and creating a tangle of avoided crossings in the butterfly potential wells. Fig. \ref{fig:RbMs} shows the PECs for larger values of $\Omega$.  As $\Omega$ increases the allowed $J$ values also increase, eliminating some PECs until for the highest nontrivial $\Omega$ value only a $^3P_2$ potential curve of $\Pi$ symmetry remains.

  \begin{figure}[t]
{\normalsize 
\includegraphics[scale =0.09]{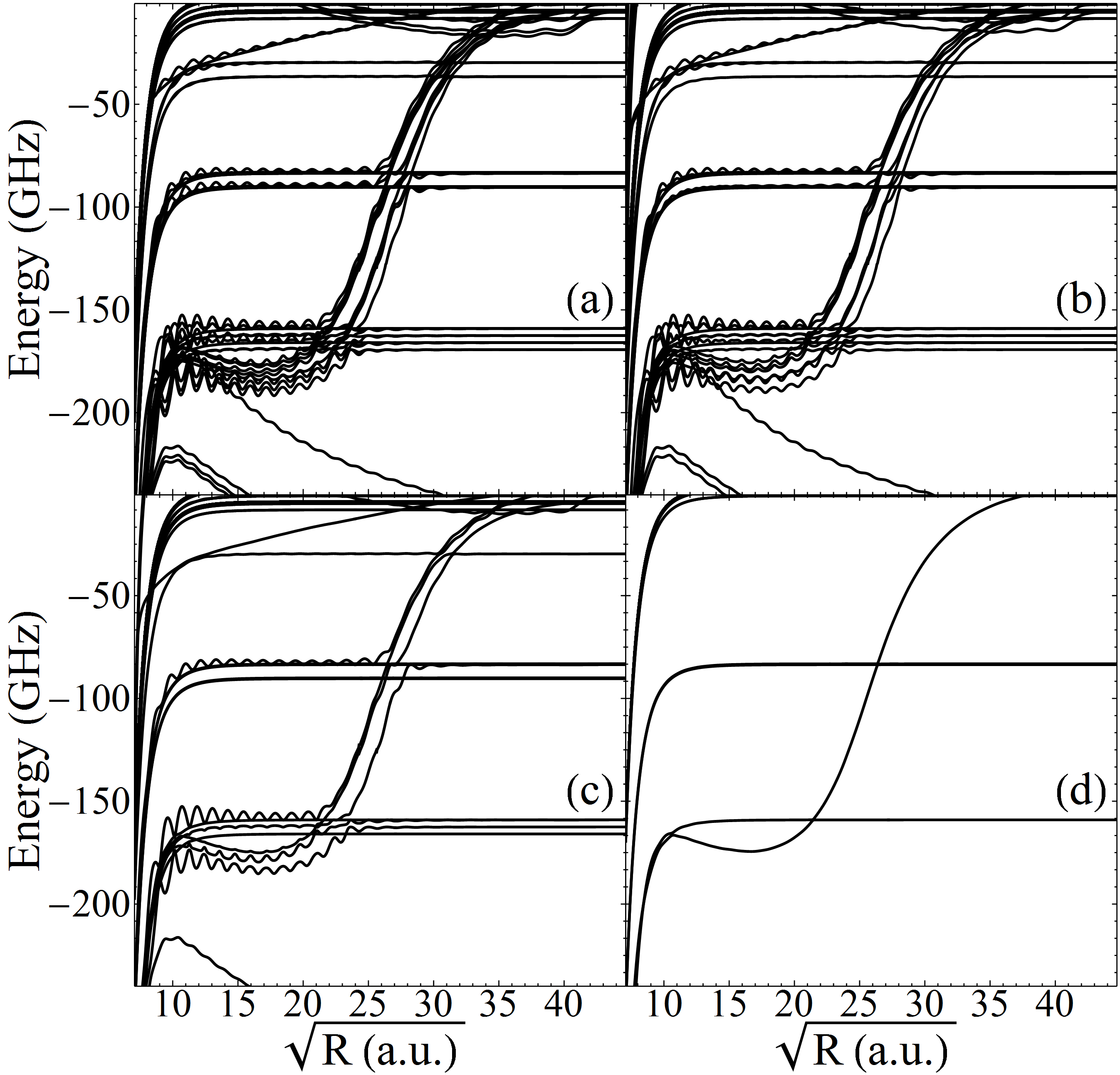}
}
\caption{PECs of Rb$_2$, for a) $\Omega = 1/2$;  b) $\Omega = 3/2$; c) $\Omega = 5/2$; d) $\Omega = 7/2$. The detuning is relative to $n_H=30$.   }
\label{fig:RbMs}
\end{figure}
\begin{figure}[tbp]
{\normalsize 
\includegraphics[scale =0.09]{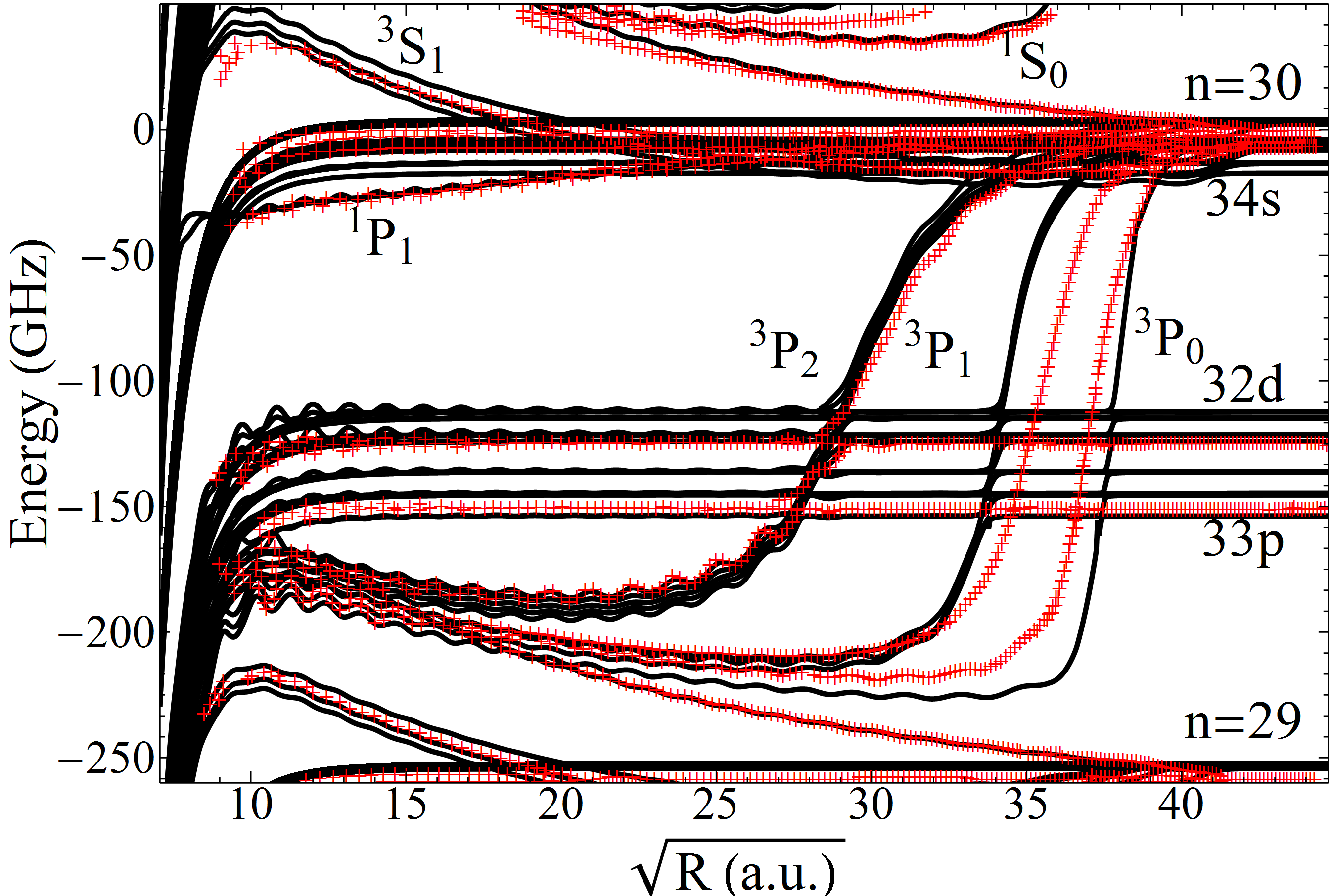}
}
\caption{PECs of Cs$_2$, including the hyperfine splitting of the ground state atom, for the projection $\Omega = 1/2$ are plotted in black. The KCF results are shown as red crosses.    The detuning is relative to $n_H=30$.}
\label{fig:CsM1n30}
\end{figure}

 Fig. \ref{fig:CsM1n30} is the same as Fig. \ref{fig:RbM1n30}, but for Cs$_2$. Again, the major features of the KCF potential curves are reproduced excellently, but several discrepancies necessitate discussion. The larger hyperfine and fine-structure splittings of Cs create significant differences in the low-$l$ asymptotes and crossings with the $^3P_J$ butterfly states. The main differences in the $^3P_J$ states are due to the modified phase shifts, since those employed here were modified to reflect direct experimental input. Differences remain, particularly in the ultra-long-range $^3P_0$ state, even when identical phase shifts are used. These discrepancies, appearing particularly at long-range and low scattering energy, are also visible in in the long-range ``trilobite'' region at the order of a few GHz. The alternative Green's function approach utilizing zero-range potentials of \cite{HamiltonGreeneSadeghpour} agrees closely with the diagonalization results presented here, suggesting that these differences stem from the finite range potential formalism of KCF.

\begin{figure}[b]
{\normalsize 
\includegraphics[scale =0.087]{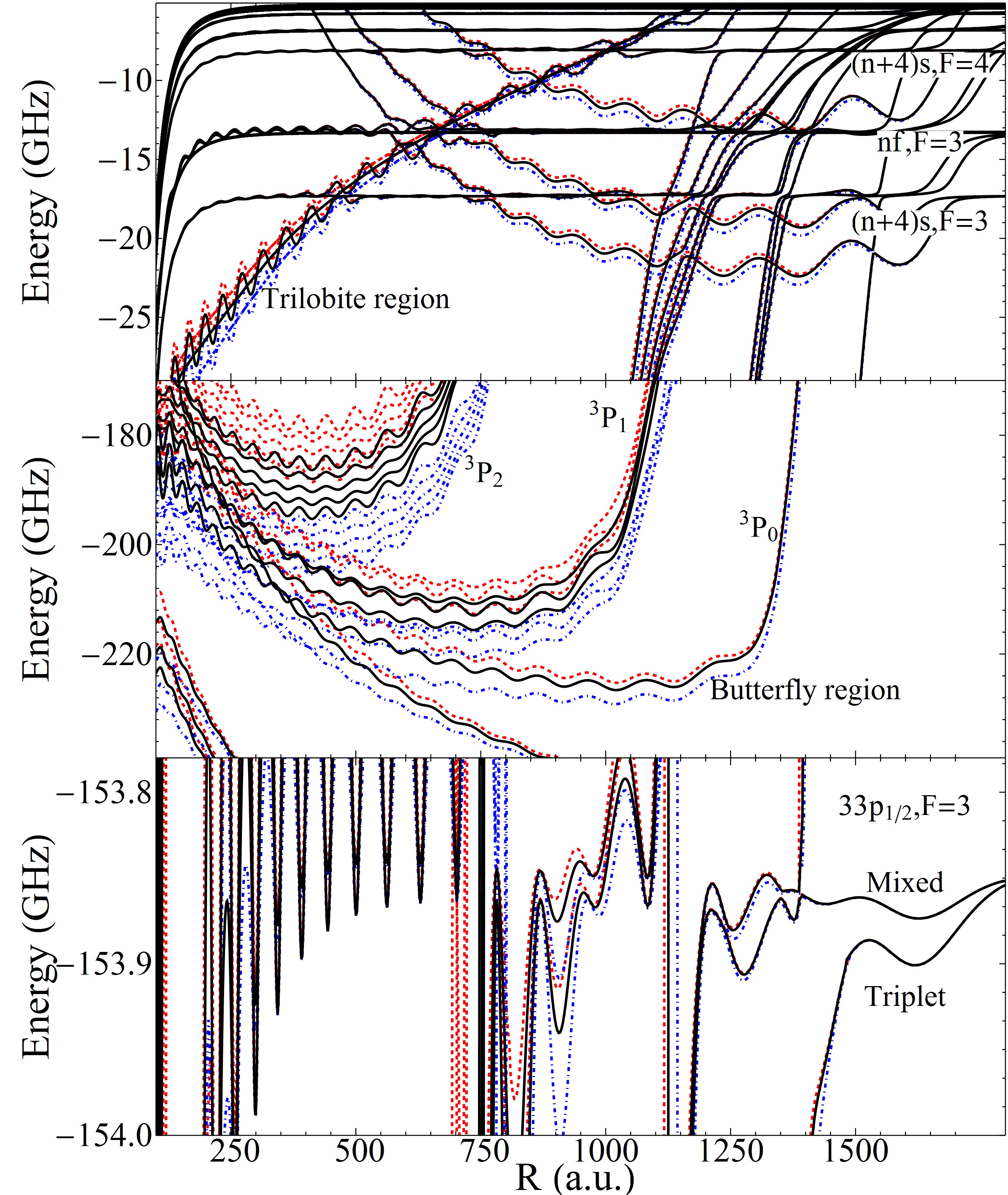}
}
\caption{ PECs of Cs$_2$, $\Omega = 1/2$, relative to $n_H=30$. The results using the $\{29,30,31\}$ basis (dot-dashed, blue), the $\{28,29,30,31\}$ basis (solid, black) and the $\{27,28,29,30,31\}$ basis (dashed,red) are plotted. Each panel shows a different regime of the PECs, showing that at long-range the calculation is quite well converged with either basis, but the short-range butterfly curves in particular vary severely with the basis size.  }
\label{fig:convergencecomparison}
\end{figure}

As a numerical test of the convergence of these results, three different basis sets ($\{n_H-q,...,n_H,n_H+1\}$, with $q = 3,2,1$) were used to calculate the PECs of Cs$_2$ in three of the most interesting regimes. These comparisons are shown in Fig. \ref{fig:convergencecomparison}. At long range the inclusion of additional manifolds {\it below} the level of interest does not contribute to the non-convergent increase in well depth seen by \cite{Fey}, but at short range these additional manifolds have a strong effect on the potential wells, repulsing them upwards. Setting $q=2$ agrees well with KCF and BK. An expanded convergence test was additionally performed for basis sets $\{n_H - q,...,n_H+p\}$. with $q=1,2,...6$ and $p=1,2$, giving an estimated uncertainty of 3GHz for the butterfly states and 5MHz for the long-range states. This uncertainty in the butterfly states applies to their absolute depths since issues with the basis size is manifested primarily as a global shift. The shape and relative depth of the individual wells is less sensitive, and the uncertainty on the relative energies of observed states is estimated to be about 0.5 GHz. 

As a final comparison, the observed butterfly states of Rb are considered in Fig. \ref{fig:Rbbutterfly}. Overlayed onto the PECs are the observed bound states (red points), whose bond lengths, extracted from rotational spectra, fix them as points in the two-dimensional energy/position plane. Additionally, the full spectrum is overlayed as horizontal lines, showing the range of energies and change in density of states as higher excited states are observed. Qualitative agreement is observed for both these comparisons, although at shorter internuclear distances the observed states are further detuned than our PECs allow. This could be due several factors: the potential wells here are very sensitive to the $^3P_J$ phase shifts; this could reflect further problems with the convergence of these PECs; or, this might signify the presence of $D$-wave scattering.    Future work is required to determine if the simple delta function potentials truly cannot be accurately converged, and if either a Green's function method or a more suitable set of basis configurations are necessary \cite{Fey}. Some likely improvements include: a varying number of basis states as a function of $R$,  an R-matrix treatment along the lines of the recent study by \cite{TaranaCurik},   or the renormalization method of \cite{DuGreene89}. Additionally, some of these problems might stem from the use of the semiclassical electron momentum;  $k(R)$ could be modified self-consistently until a converged result is attained.

\end{section}
\begin{section}{Discussion}

The elements investigated in a photoassociation process determine many key properties of the Rydberg molecules. The prominent differences between the two alkali atoms considered here are their quantum defects and $^3P_J$ scattering properties. The top panel of Fig. \ref{fig:convergencecomparison} shows the PECs in the Cs$_2$ ``trilobite'' region near the $n_H = 30$ manifold. The near-degeneracy between the $(n+4)s$ states and this manifold allows two-photon excitation of the trilobite molecule \cite{Tallant,trilobite}; this is not reasonable in Rb since the trilobite state admixes almost exclusively high-$l$ states.

Likewise, the positions of the $^3P_J$ shape resonances and their energy dependences strongly change the butterfly potential wells. The $^3P_0$ resonance in cesium occurs at such a low electronic energy that the associated PECs cross the low-l states at very large internuclear distances, destabilizing the longest-range states to a greater degree than in Rb (Fig. (\ref{fig:Rbbutterfly}) displays rubidium's $np$ and butterfly PECs). The butterfly states of Rb possess significant $p$-character, making a single-photon excitation through this admixture possible;  the butterfly states of Cs are much further detuned from the $np$ asymptotes (e.g. see Fig. \ref{fig:CsM1n30}), and possess less $p$ character. Additionally, the much larger $^3P_J$ splittings in Cs greatly spread the butterfly wells, limiting the number of avoided crossings.

\begin{figure}[t]
{\normalsize 
\includegraphics[scale =0.045]{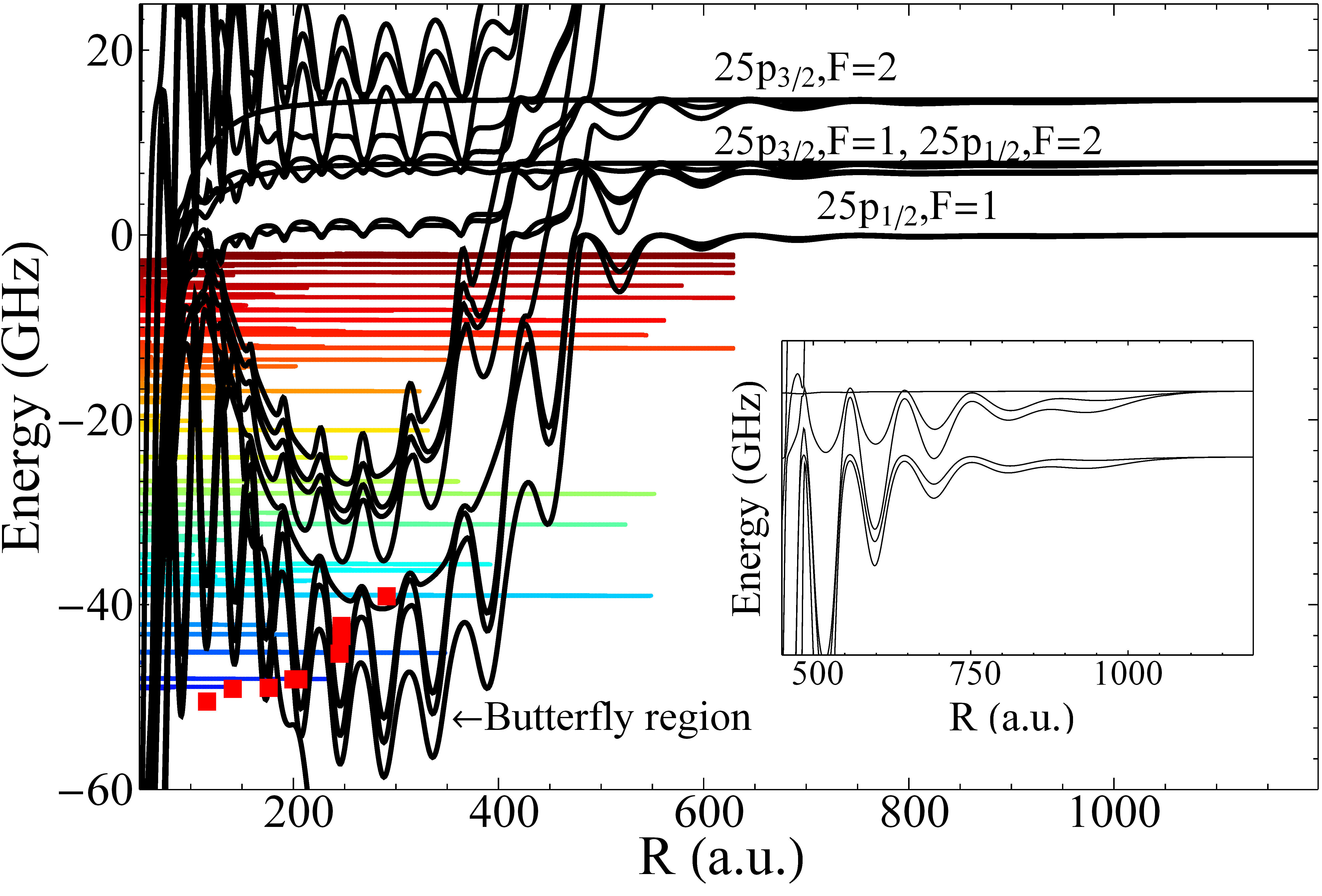}
}
\caption{$\Omega = 1/2$ Rb$_2$ PECs (black/solid) near the $25p$ Rydberg states, which descend into the butterfly potential wells at short internuclear distances. Zero energy is set to the 25p$_{1/2},F=1$ asymptote. The bound states whose PEDMs were characterized in Ref. \cite{butterfly} are plotted as red squares, while the observed spectrum of that experiment is overlayed. The color scheme matches that of Ref. \cite{butterfly}, and has no meaning but to guide the eye. The 25p$_{3/2},F=1$ and 25p$_{1/2},F=2$ potential wells are highlighted in the inset, since for this Rydberg level the interplay between the fine and hyperfine states makes these states nearly degenerate.  }
\label{fig:Rbbutterfly}
\end{figure}
The interplay between different fine and hyperfine splittings can also be used to engineer Rydberg molecules with specific spin characters, and notably can be tuned via the principal quantum number to induce spin flips in the perturbing atom or to strongly entangle the nuclear spin of the perturber with the electronic spin of the Rydberg atom \cite{Niederprum}. The PECs for these states are highlighted in Fig. (\ref{fig:Rbbutterfly}). In particular, the near-degeneracy of $25p_{3/2},F=1$ and $25p_{1/2},F=2$ states strongly mixes their spin character; this degeneracy can be varied over a range of quantum numbers from 24-29.  Similar degeneracies are found in the $np$ states of Cs, for $n = 31-35$ \cite{Markson}, or also in the Cs $nd$ states for $n = 21-25$. The myriad differences between these two alkali species provide a wide range of parameters influencing the properties of the Rydberg molecules, and future work could investigate how the impact of different properties of other alkali atoms such as Li, Na \cite{LiNaNorcross}, K,  or Fr \cite{RbCsFr} in their respective long-range Rydberg molecules. Other interesting opportunities involve studies of heteronuclear Rydberg molecules: for example, an excited Cs atom bound to a ground state Rb atom would take advantage of the favorable near-degeneracy between the $(n+4)s$ and $n,l>3$ energies without the added complications of the large $^3P_J$ splitting of the e-Cs scattering resonances.  Recent work has demonstrated that an even wider diversity of excitation pathways, final molecular states, and decay channels can be found in {\it non-alkali} atoms, due to their complex multichannel behavior \cite{MattPaper,SrPaper1,SrPaper2}.  One class of such multichannel atoms, the alkaline-earths, provide additional simplifications, as they lack hyperfine structure for the most common isotopes and, for those heavier than Mg, $P$-wave shape resonances \cite{Bartschat}.

  \end{section}
\begin{section}{Multipole moments}

The state mixing induced by the perturber creates large permanent electric dipole moments (PEDM) in these molecules. This even occurs in the weakly perturbed low-$l$ states due to small admixtures of trilobite or butterfly states \cite{PfauSci}. Since the PEDMs of both trilobite and butterfly molecules have been observed in recent experiments \cite{trilobite,butterfly}, new interest in the application of these molecules in dipolar gases and ultracold chemistry has been sparked. The higher multipole moments of these molecules are of interest for detailed calculations of the inter-molecular interactions. 

The multipole moments of the $i$th electronic configuration are 
$
d_{\nu,i}^{k,q} =\bra{i}T^k_q\ket{i},
$
where the multipole moments from classical electrostatics \cite{Jackson} are promoted to quantum-mechanical operators: 
\be
T^k_q = -r^k\sqrt{\frac{4\pi}{2k+1}}Y_{k,q}(\hat r).
\ee
We first generalize the dipole moments of KCF, derived in the absence of spin and for purely hydrogenic states, are generalized to all orders in the multipole expansion.  The $\Sigma$ trilobite ($\Psi_{00}(\vec R, \vec r)$) and $\Sigma,\Pi$ butterfly ($\Psi_{10}(\vec R,\vec r),\Psi_{1\pm 1}(\vec R,\vec r)$, respectively) states are well described by the electronic wave functions \cite{KhuskivadzeChibisovFabrikant}\begin{align}
\Psi_{LM_L}(\vec R,\vec r) &= \frac{\sum_{l}Q_{LM_L}^{nl}(R)Y_{lM_L}(\hat r)r^{-1}f_{nl}(r)}{\sqrt{\sum_l|Q_{LM_L}^{nl}(R)|^2}},
\end{align} where the $j$-dependence of the $Q$ functions (see Eq. (\ref{eqn:Qfuncs})) is removed, the sum over $l$ spans from $l_{min}\simeq 2$ to $n-1$, and $r^{-1}f_{nl}(r)Y_{lm}(\hat r)$ is the Rydberg wave function. Using these approximate forms, which ignore couplings to other $n$ manifolds and assume vanishing quantum defects, the multipole moments are 
\begin{align}
\langle T_q^k\rangle &=\bra{\Psi_{L'M'}}T_q^k\ket{\Psi_{L'M'}}\\ &= \sum_{l,l'}\frac{Q_{L'M'}^{nl}(R)Q_{L'M'}^{nl'}(R)}{\sum_l|Q_{L'M'}^{nl}(R)|^2} \bra{nlM'}T_q^k\ket{nl'M'}.\nonumber
\end{align}
The matrix element separates into a radial integral, $R_{nl}^{nl'}(L) = \int dr f_{nl}(r)r^Lf_{nl'}(r)$, and an angular integral which is expressed as a reduced matrix element through the Wigner Eckart theorem:
\begin{align}
\label{eqn:wignereckart}
\langle T_q^k\rangle&= \sum_{l,l'}\frac{Q_{L'M'}^{nl}(R)Q_{L'M'}^{nl'}(R)}{\sum_l|Q_{L'M'}^{nl}(R)|^2} R_{nl}^{nl'}(L)\\&\times\frac{C_{l'M',kq}^{lM'}}{\sqrt{2l'+1}}\bra{l|}T^k\ket{|l'},\nonumber
\end{align}
where
\be
\label{eqn:reducedmatel}
\bra{l|}T^k\ket{|l'}= (2l'+1)\begin{pmatrix}l & l' & k\\ 0 & 0 & 0 \end{pmatrix}(-1)^{k-l'}.
\ee
\begin{figure}[t]
{\normalsize 
\includegraphics[scale =0.08]{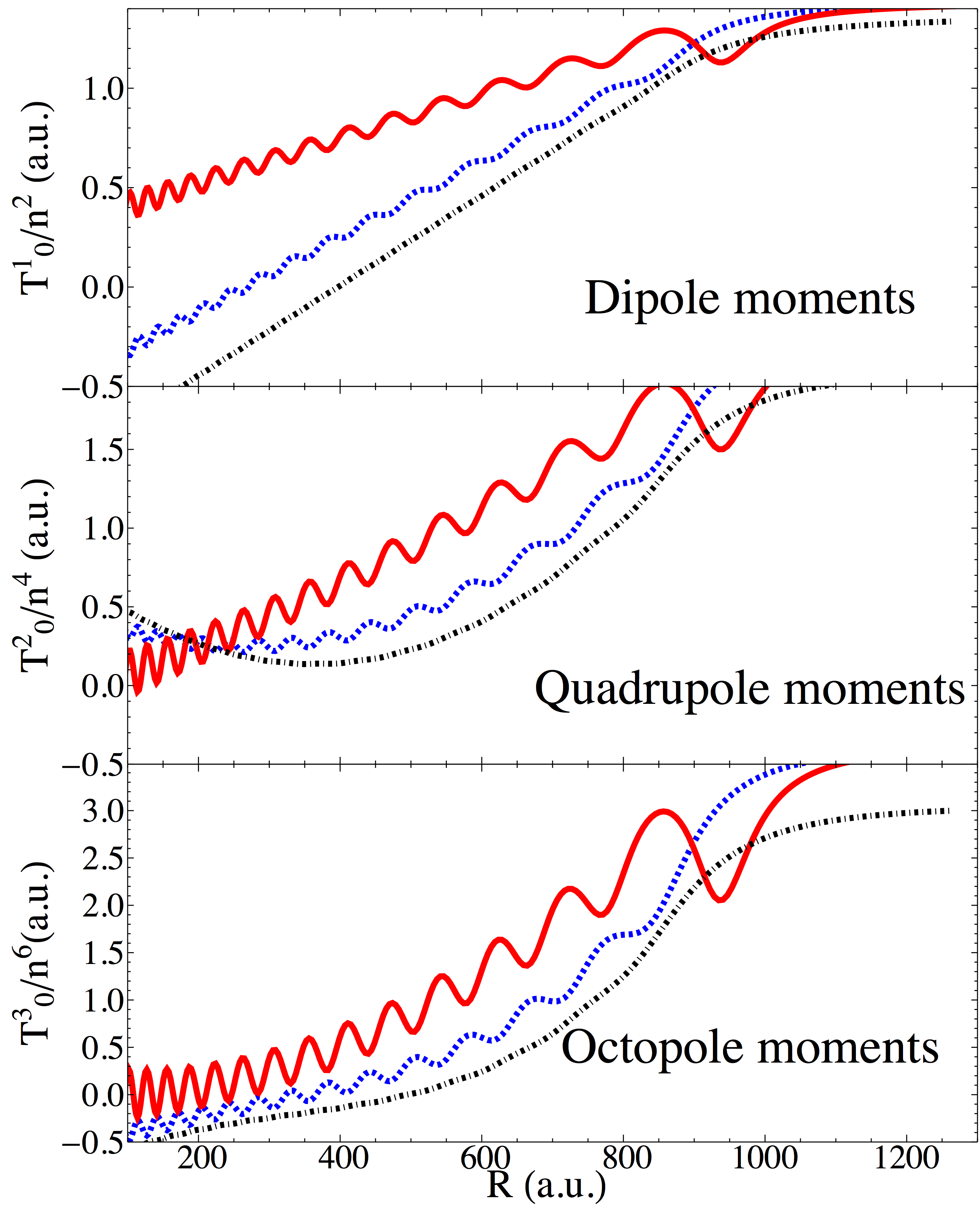}
}
\caption{ Analytic (valid only for hydrogenic states) dipole, quadrupole, and octupole moments for $n =23$, using Eq. (\ref{eqn:analyticdipole}). The trilobite (blue,dashed) and $\Sigma$ butterfly state (red,solid) oscillate as a function of $R$, while the the $\Pi$ butterfly state (black,dot-dashed) is non-oscillatory; this behavior matches the PECs. }
\label{fig:multipoles}
\end{figure}
Eqs. \ref{eqn:wignereckart} and \ref{eqn:reducedmatel} lead to the result
\begin{align}
\label{eqn:analyticdipole}
\langle T_q^k\rangle &= \sum_{l,l'}\frac{Q_{L'M'}^{nl}(R)Q_{L'M'}^{nl'}(R)}{\sum_l|Q_{L'M'}^{nl}(R)|^2}  R_{nl}^{nl'}(k)\\& \times C_{l'M',kq}^{lM'}(-1)^{k-l'}\sqrt{(2l'+1)}\begin{pmatrix}l & l' & k\\ 0 & 0 & 0 \end{pmatrix}.\nonumber
\end{align}
The Clebsch-Gordan coefficient causes any term with $M'\ne 0$ to vanish, reflecting the cylindrical symmetry.  The $L = 1$ moments agree exactly with KCF. These multipole moments scale in size as $n^{2L}$, and are displayed in Fig. \ref{fig:multipoles} up to the octupole moments.

Within the full spin model, the multipole moments are derived similarly, but using the numerically calculated eigenstates, $\ket{s} = \sum_{k}a_{sk}\ket{k}$, where $\ket{s}$ is an electronic eigenstate, $k$ is a composite quantum number $k=\{n(ls_1)jm_jm_2m_i\}$, and $a_{sk}$ is the eigenvector corresponding to the $s$th eigenstate.  The multipole moments are then
\begin{align}
\label{eqn:fullmultipoles}
&\bra{s}T_q^k\ket{s} = \sum_{k,k'}\Bigg[a_{sk}a_{sk'}\delta_{m_2,m_2'}\delta_{m_i,m_i'}\delta_{m_j,m_j'}R_{nlj}^{n'l'j}(L)\nonumber\\& \times(-1)^{s_1+j'+l-l'}\sqrt{(2j+1)(2l+1)(2l'+1)}C_{j'm_j',kq}^{jm_j}\nonumber\\&\times\begin{pmatrix}l & l' & k\\ 0 & 0 & 0 \end{pmatrix}\begin{Bmatrix}l' & s_1 & j'\\ j & k & l\end{Bmatrix}\Bigg].
\end{align}

\begin{figure}[b]
{\normalsize 
\includegraphics[scale =0.35]{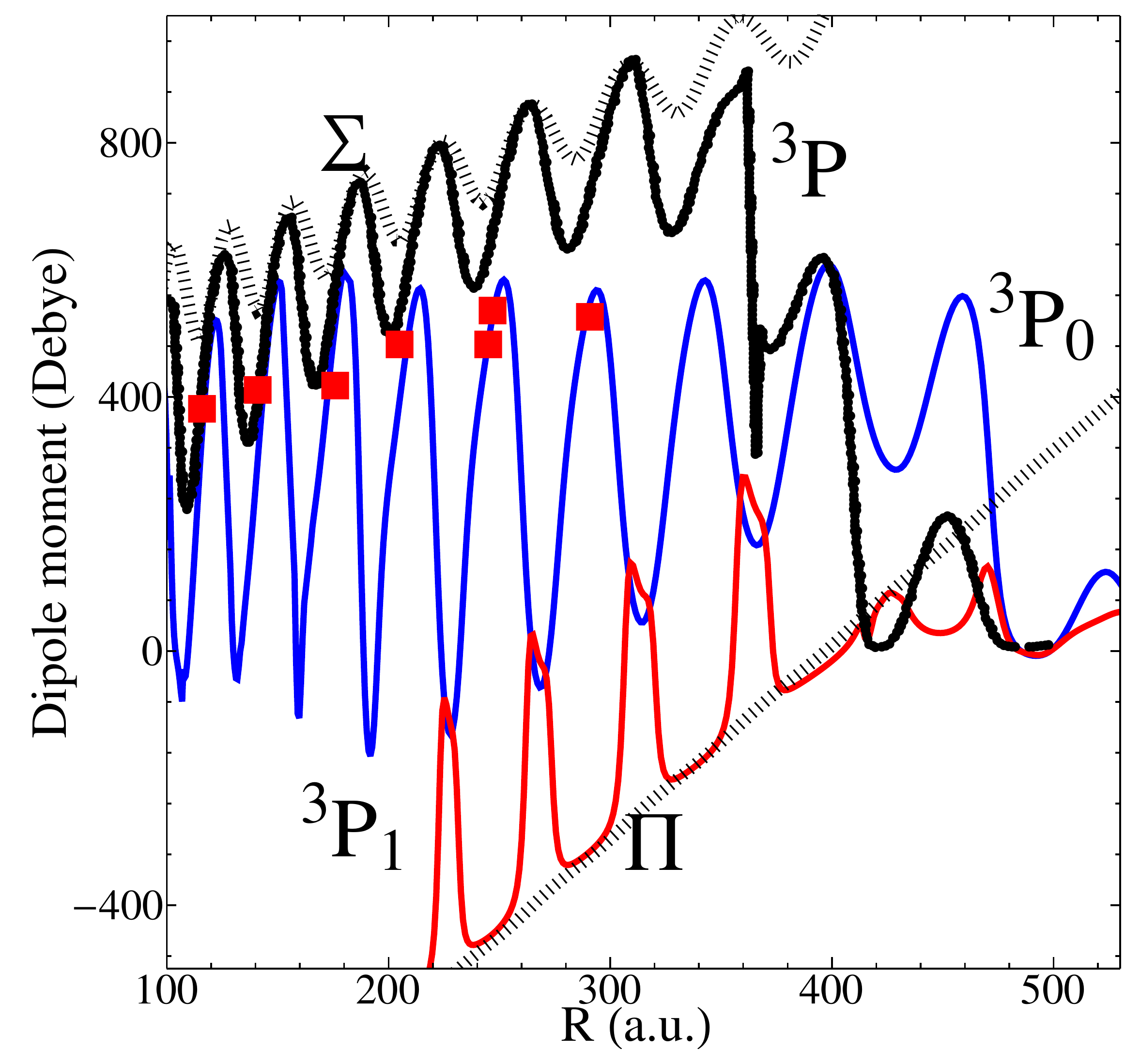}
}
\caption{(color online) Analytic PEDMs (black, dashed), PEDMs ignoring the $^3P_J$ splitting (black, solid, labeled $^3P$), and PEDMs from the full spin model for electronic states dominated by $^3P_0$ scattering (blue,solid), and $^3P_1$ scattering (red,solid), are plotted. The red squares are placed at the observed bond lengths and PEDMs \cite{butterfly}.  The $^3P_0$ and $^3P_1$ PEDMs correspond to states of mixed $M_L$, although the mixing is quite weak for $^3P_1$ scattering and the analytic and exact results agree more closely.  The $^3P_2$ case is not shown, for simplicity.}
\label{fig:dipoles}
\end{figure}
Several PEDMs are plotted in Fig. \ref{fig:dipoles}: those corresponding to $^3P_0$ and $^3P_1$ states, the analytic curves for $\Sigma$ and $\Pi$ symmetries of Eq. \ref{eqn:analyticdipole}, the butterfly curve neglecting $^3P_J$ splitting used in Ref. \cite{butterfly}, and the experimentally observed values. The observable PEDMs are obtained from the theoretical curves by averaging over the vibrational wave functions of the relevant states. The $^3P_0$ PEDM is noticeably smaller and oscillates more dramatically than the $\Sigma$ and $^3P$ curves. The maxima in this curve are correlated with the positions of bound states in the relevant potential wells. At large $R>400$  these PECs connect adiabatically with the $np$ states, which explains the rapid decrease in the PEDMs as $R$ increases.

The reduced strength of the $^3P_0$ PEDM relative to the results neglecting $J$-dependence (the $^3P$ curve) stems from the $M_L$-mixing caused by the SO splitting of the electron-perturber interaction. The PEDMs extracted from pendular state measurements are systematically smaller (by $\sim$25\%) than predicted by the $^3P$ curve (solid black), which follows the approximate $\Sigma$ curve quite closely \cite{butterfly}. The full theory explains this systematic difference: the $M_L = 0$ states focus the electronic wave function near the perturber, while $|M_L| = 1$ states maximize the wave function closer to the Rydberg core ion; their mixing places the mean value of the electron's position closer to the positively-charged core and reduces the PEDM. Examination of Fig. \ref{fig:multipoles}a reveals that any mixture of $M_L$ in this region of internuclear distances mixes negative and positive PEDMs, reducing the total strength. Quantitative agreement is seen between the experimental PEDMs and the the theoretical curves they lie directly on at the bond lengths extracted from the experiment, which also agree with the potential minima predicted by the theory. The $^3P$ prediction does not even overlap most experimental points. This is evidence that even though the relatively small e-Rb $^3P_J$ scattering splittings do not dramatically shift the PECs, these splittings do have significant impact on observables such as the PEDMs. For Cs, this effect will be even greater. Further insight into this spin mixing is given by considering the $^3P_1$ curve, which is predominantly a $\Pi$ symmetry state except for hyperfine-induced mixing, which occurs near avoided crossings of the potential curves. Fig. \ref{fig:dipoles} shows that this PEDM lies on the straight line predicted by the approximate $\Pi$ curve, except for deviations located at avoided crossings in the relevant potential curves.

\end{section}

\begin{section}{Conclusions}
A full theoretical model has been presented here which accurately includes all relevant relativistic effects. This effort serves as a foundation for future experimental efforts requiring the most complete theoretical picture, and provides a basis for future theoretical work studying new systems or novel applications of these exotic molecules.  Furthermore, this development will soon be combined with accurate approaches for calculating the binding energies and line strengths of these bound states in order to quantitatively assess the agreement with experiment \cite{paper2}. The recent observation of butterfly molecules is a promising step towards the routine preparation of enormous dipolar ultracold molecules, which will be new paradigms of controllability at scales far beyond the state of the art. The results presented here help to better understand the character of these molecules, as well as their binding energies and PEDMs. The prospects of forming these butterfly molecules in Cs will perhaps be more challenging since the $p$ character of the butterfly state is much smaller, but the huge separation between $^3P_J$ potential curves greatly enlarges the range of internuclear distances and PEDMs accessible in these molecules. The improved description of the nearly-degenerate high-$l$ manifold with the very close $(n+4)s$ state given here lends a more complete theoretical description of this state that should encourage further exploration of the trilobite state in Cs.

Ultracold chemical processes related to these two systems (where $X$ can be either Rb or Cs) are also of current interest, namely $l$-changing collisions leading to the formation of X$^+$ ions, or the formation of $X_2^+$ molecular ions. The former process occurs due to nonadiabatic processes creating pathways from the initial state to asymptotic regions correlated with other angular momentum states, or even to the high-$l$ manifold via couplings with the trilobite state. The latter process occurs when the neutral atom tunnels inwards out of the potential well in which the molecule is bound towards the Rydberg core. This process is facilitated by the $^3P_J$ potential curves, which descend so steeply from the low-$l$ asymptotic states that the neutral atom is accelerated to the short-range region where ultracold chemistry can occur. This has been studied in some detail in Rb, and some recent experimental investigations along similar lines focusing on the decay channels has been reported for Cs as well \cite{UltracoldChem,CsReview}. 
\end{section}
 
\begin{acknowledgements}
This work is supported in part by the National Science 
Foundation under Grant No. NSF PHY-1607180 and under Grant No. NSF PHY11-25915. We gratefully acknowledge 
J. P\'{e}rez-R\'{i}os and P. Giannakeas for many enlightening discussions and insight. MT Eiles is grateful for the hospitality of the Hamburg CUI and for many fruitful discussions with participants of the ``From few- to many-body physics in cold atomic quantum matter'' workshop during the early stages of this manuscript. 
\end{acknowledgements}
\appendix
\begin{section}{Appendix A: Derivation of projection operator form of the pseudopotential}

 The Fermi $(L=0)$ pseudopotential along with the Omont generalization to $P$-wave interactions are
\begin{align*}
V_P(\vec r,\vec R) &= 2\pi\sum_{L=0}^1(2L+1)a(L,k)\cev{\nabla}{}^L \cdot\delta^3(\vec  r - \vec R)\vec\nabla^L.
\end{align*}
 $a(L,k)$ is the energy-dependent scattering length(volume) for $L = 0(1)$, and $L$ is the electron's angular momentum in the coordinate system defined by the perturber, where the electron is located at $\vec X = \vec r - \vec R$ . The gradient terms act on the bra or ket, following the direction of the vector arrow. Spin-dependence is introduced via projectors onto singlet or triplet states and the inclusion of spin-dependent scattering parameters:
\begin{align*}
V_P &= 2\pi\sum_{S,M_S}\sum_{S',M_S'}\ket{SM_S}\bra{SM_S}\\&\times\sum_L(2L+1) a(SL,k)\cev\nabla{}^L\cdot\delta^3(\vec X)\vec\nabla^L\ket{S'M_S'}\bra{S'M_S'}.
\end{align*}
We desire a form of the pseudopotential in terms of projections onto states of total $J$, $V_P \propto\sum_J\ket{(LS)J\Omega}A_{LSJ}\bra{(LS)J\Omega}$. This is accomplished by projecting the above equation onto states of $L$, leaving only an integration over the radial part in the eventual construction of the matrix elements:
\begin{widetext}
\begin{align*}
V_P &= 2\pi\sum_{\substack{LM_L\\L'M_L'}}\sum_{S,M_S}\sum_{L''}(2L''+1)a(SL'',k)\delta_{S,S'}\ket{LM_L,SM_S}\int Y^*_{LM_L}(\hat X)\cev\nabla{}^{L''}\cdot \delta^3(\vec X)\vec\nabla^{L''}Y_{L'M_L'}(\hat X)d\hat X\bra{L'M_L',SM_S}.
\end{align*}
This expression is diagonal in $S$ and also in $L$, since the basis states are expanded in powers of $X^L$; if the exponent on the gradient operator does not match the exponent of $X$, these terms vanish. This leaves:
\begin{align*}
V_P&=2\pi\sum_{LM_L,M_L'}\sum_{S,M_S}(2L+1)a(SL,k)\ket{LM_L,SM_S} \int Y^*_{LM_L}(\hat X)\cev\nabla{}^{L}\cdot \delta^3(\vec X)\vec\nabla^{L}Y_{L'M_L'}(\hat X)d\hat X\bra{LM_L',SM_S}.
\end{align*}
The integration over $\hat X$ can now be performed. Explicitly, for $L=0$:
\begin{align*}
 \int Y^*_{00}(\hat X) \delta^3(\vec X)Y_{00}(\hat X)d\hat X&=\frac{\delta(X)}{X^2}Y_{00}(0,0)Y_{00}(0,0).
\end{align*}
And, for $L=1$:
\begin{align*}
 \int Y^*_{1M_L}(\hat X)\cev\nabla\cdot \delta^3(\vec X)\vec\nabla Y_{1M_L'}(\hat X)d\hat X=\frac{\delta(X)}{X^2}\left(\partial_X'\partial_X Y_{1M_L}(0,0)Y_{1M_L'}(0,0) +\frac{1}{X^2}\frac{(2L+1)(L+1)L}{8\pi}\delta_{M_L,M_L'}\delta_{|M_L|,1} \right).
\end{align*}
Here $\partial_X'\partial_X$ is the radially-dependent term of the dot product of the two gradient operators, where $\partial_X'$ acts to the left. Since the analysis in the text only considers functions linear in $X$ for $L=1$, the derivative term can be effectively replaced by a $X^{-2}$ factor to give the following compact form for the full pseudopotential:
\begin{align*}
V_P&=2\pi\sum_{LM_L,M_L'}\sum_{SM_S}\ket{LM_L,SM_S} \frac{(2L+1)^2}{4\pi}a(SL,k)\frac{\delta(X)}{X^{2(L+1)}}\delta_{M_L,M_L'}\bra{LM_L',SM_S}.
\end{align*}
The angular momenta may now be coupled and summed over $M_L'$:
\begin{align*}
V_P&= 2\pi\sum_{LM_L}\sum_{SM_S}\sum_{J\Omega,J'\Omega'}\ket{(LS)J\Omega} C_{LM_L,SM_S}^{J\Omega} \frac{(2L+1)^2}{4\pi}a(SLJ,k)\frac{\delta(X)}{X^{2(L+1)}}C_{LM_L',SM_S}^{J'\Omega'}\bra{(LS)J'\Omega'}
\end{align*}
Summation over $M_L$ and $M_S$ replaces the product of Clebsch-Gordan coefficients with $\delta_{JJ'}\delta_{\Omega\Omega'}$, along with the triangularity condition relating the possible values of $L$ and $S$ to the allowed values of $J$. Finally,
\begin{align}
\label{asd}
V_P&= 2\pi\sum_{(L,S)J\Omega}\ket{(LS)J\Omega}\frac{(2L+1)^2}{4\pi}a(SLJ,k)\frac{\delta(X)}{X^{2(L+1)}} \bra{(LS)J\Omega},L\le 1
\end{align}
\end{widetext}
This pseudopotential form, with the angular dependence situated in the projectors, is the desired form to incorporate the $J$-dependent scattering parameters correctly. 
    \end{section}
    \begin{section}{Appendix B: Further comments on the effects of $^3P_J$ scattering}
  To highlight the impact of the $^3P_J$ splitting effect, it is isolated by ignoring the fine structure of the Rydberg atom and the hyperfine structure of the ground state atom. In this case the matrix elements of the Fermi/Omont pseudopotential are given in the total spin basis, $\ket{nlmSM_S}$, where $\ket{nlm}$ is the wave function of the Rydberg electron.  For definiteness, only the $L=1,S=1,\Omega=0$ matrix elements are considered:
 \begin{align}
 \label{eqn:appendixB2}
  &\bra{nlmSM_S}\hat V\ket{n'l'm'S'M_S'}\\ &= \sum_{J}6\pi a(11J,k)\delta_{m,-M_S}\delta_{m',-M_S'}\nonumber\\&\times C_{1m,1-m}^{J0}C_{1m',1-m'}^{J0}Q_{1m}^{nl}(R)Q_{1m'}^{n'l'}(R)\delta_{S,S'}\nonumber.
 \end{align}
If the scattering volume's $J$-dependence is neglected, the $J$-summation can be performed over the two remaining Clebsch-Gordan coefficients, yielding a diagonal matrix in $m,m'$, as expected. In contrast, Eq. (11) of \cite{Markson} gives the same result as eq. \ref{eqn:appendixB2} multipled by $\delta_{mm'}\delta_{M_SM_S'}$.
 
 Fig. \ref{fig:matels} displays the matrix elements of the $^3P_J$ scattering potential within a restricted Hilbert space, with fixed $n,n',l,l'$. Since $\Omega=0$, only states with opposite $m,M_S$ are nonzero. The size of the non-diagonal elements reflects the mixing of $m$ values.  As Fig. \ref{fig:matels2} illustrates, these off-diagonal elements are truly essential in capturing the physics of this process, as they are needed to obtain three distinct eigenvalues out of different linear combinations of states of different $m$. If only the diagonal elements are included, the eigenvalues labeled 2 and 3 are degenerate, and only two butterfly potential wells develop. 
 \begin{figure}[tbp]
{\normalsize 
\includegraphics[scale =0.6]{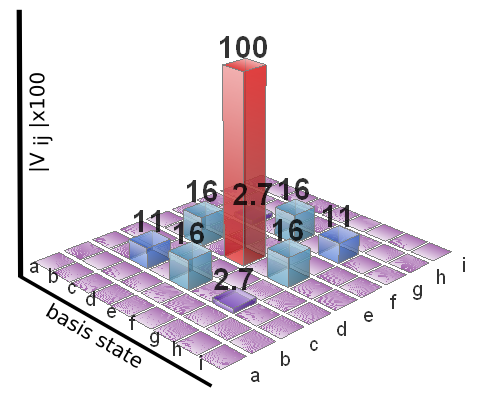}
}
\caption{Absolute (normalized so that the largest is 100) values of the elements of the scattering matrix $V_{ij}=\bra{nlS,mM_S}\hat V\ket{nlS,m'M_S'}$ at $R = 700$ and with $n=30$,$l=10$, and $S=1$, for Cs. The basis states are labeled by $\ket{mM_S}$; labels $a,b,...i$ correspond to $\ket{11},\ket{01},\ket{-11},\ket{10},\ket{00},\ket{-10},\ket{1-1},\ket{0-1},\ket{-1-1}$, respectively. Ref. \cite{Markson} gives only the diagonal elements. }
\label{fig:matels}
\end{figure}
 \begin{figure}[tbp]
{\normalsize 
\includegraphics[scale =0.27]{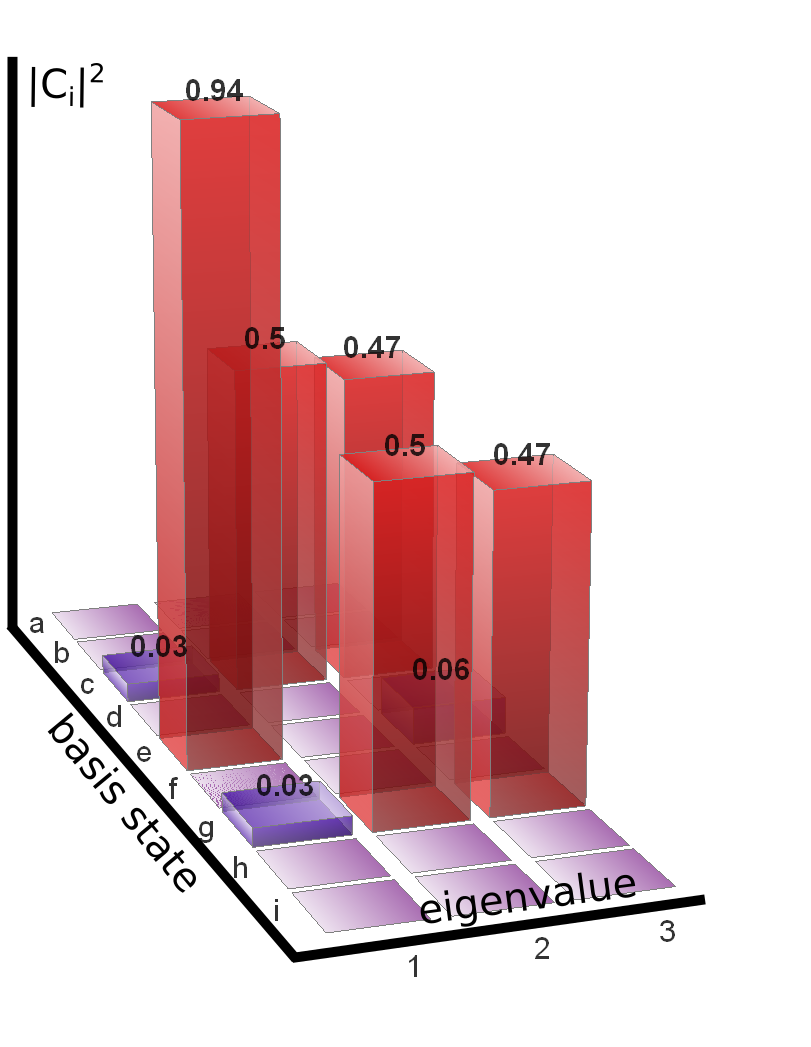}
}
\vspace{-20pt}
\caption{Absolute squares of the normalized eigenvector components, $|c_i|^2$, for the three non-zero eigenvalues. $m,m'$ are mixed for $^3P_0$ and $^3P_2$, while the $^3P_1$ scattering state has no $m=0$ component. The basis state labels are given in the Fig. \ref{fig:matels} caption.  }
\label{fig:matels2}
\end{figure}
    \end{section}
        \begin{section}{Appendix C: Rb and Cs electron scattering phase shifts}
        The phase shifts used in this work are plotted in Fig. \ref{fig:phases}, which shows the small shifts applied to the data from KCF so that the energies where the phase shift varied most rapidly corresponded to the experimentally observed resonance positions. This involved shifts of approximately 1 meV. The Rb phase shifts were not modified from those calculated in KCF. 
        \begin{figure}[tbp]
{\normalsize 
\includegraphics[scale =0.26]{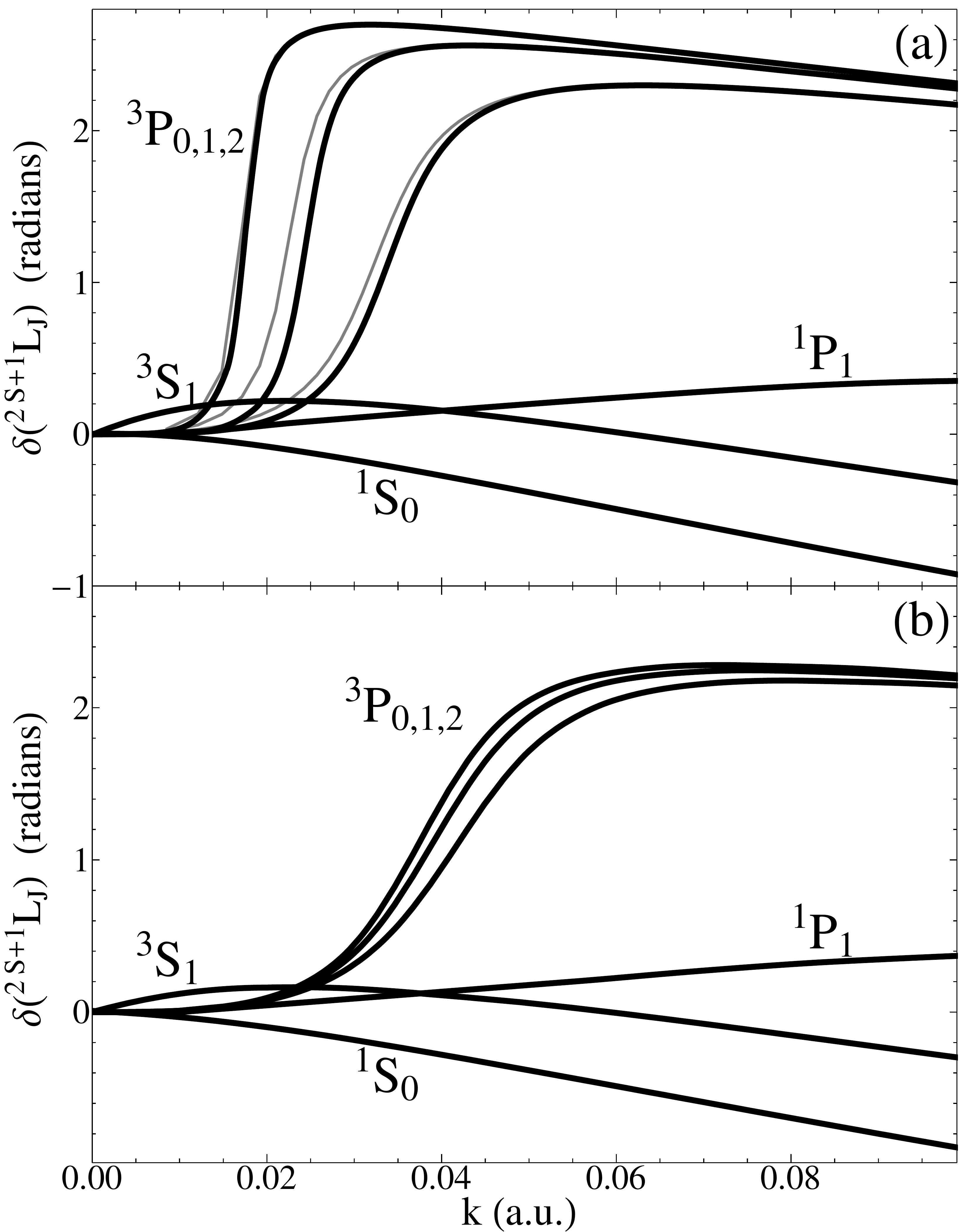}
}
\caption{Scattering phase shifts for Cs (a) and Rb (b), extracted from KCF \cite{KhuskivadzeChibisovFabrikant}. In panel a the unshifted phases are shown as faint curves; the thick curves were shifted slightly to better reflect experimentally observed resonance positions. }
\label{fig:phases}
\end{figure}
        \end{section}

\end{document}